\documentclass[runningheads]{llncs}
\usepackage{graphicx}
\usepackage{amsmath}
\begin{document}
\title{Evaluation of convolutional neural networks using a large multi-subject P300 dataset\thanks{Supported by organization x.}}
%
%
\author{Luk\' a\v s Va\v reka\inst{1}\orcidID{0000-0002-5998-3676}}
\authorrunning{F. Author et al.}
%
\institute{NTIS - New Technologies for the Information Society, Faculty of Applied Sciences, University of West Bohemia, Univerzitni 8, 306 14 Pilsen, Czech Republic \\ \email{lvareka@ntis.zcu.cz}}
\maketitle              
\begin{abstract}
Deep neural networks (DNN) have been studied in various machine learning areas. For example, event-related potential (ERP) signal classification is a highly complex task potentially suitable for DNN as signal-to-noise ratio is low, and underlying spatial and temporal patterns display a large intra- and intersubject variability. Convolutional neural networks (CNN) have been compared with baseline traditional models, i.e. linear discriminant analysis (LDA) and support vector machines (SVM) for single trial classification using a large multi-subject publicly available P300 dataset of school-age children (138~males and 112~females). For single trial classification, classification accuracy stayed between 62~\% and 64~\% for all tested classification models. When applying the trained classification models to averaged trials, accuracy increased to 76 - 79~\% without significant differences among classification models. CNN did not prove superior to baseline for the tested dataset. Comparison with related literature, limitations and future directions are discussed.

\end{abstract}
\section{Introduction}

In recent years, both fundamental and applied research in deep learning has rapidly developed. In image processing and natural language processing, it has led to significantly better classification rates than previous state-of-the-art algorithms~\cite{SIG-039}. Therefore, there has been a growing interest in applying deep neural networks (DNNs) to various fields of applied research. Such an effort can also be seen in electroencephalographic (EEG) data processing and classification. A well-known application of EEG classification is a brain-computer interface (BCI)~\cite{McFarland:2011:BIC:1941487.1941506} which allows immobile persons to operate devices only by decoding their intent from EEG signal without any need for muscle involvement. A significant challenge in BCI systems is to recognize the intention of the user correctly since the brain components of interest often have a significantly lower amplitude than random EEG signal~\cite{McFarland:2011:BIC:1941487.1941506}. 

DNNs often do not require costly feature engineering, and thus could lead to more universal and reliable EEG classification. However, recent review of the field reached a conclusion that so far, these benefits have not been convincingly presented in the literature~\cite{Lotte_2018}. Many studies did not compare the studied DNN to state-of-the-art BCI methods or performed biased comparisons, with either suboptimal parameters for the state-of-the-art competitors or with unjustified choices of parameters for the DNN~\cite{Lotte_2018}. Similar conclusion has been reached in another review of DNN and EEG~\cite{DBLP:journals/corr/abs-1901-05498}. Many related papers suffer from poor reproducibility: a majority of papers would be hard or impossible to reproduce given the unavailability of their data and code~\cite{DBLP:journals/corr/abs-1901-05498}. Moreover, one of the drawbacks of DNNs is having to collect a large training dataset. Typical BCI datasets have very small numbers of training examples, since BCI users cannot be asked to perform thousands of mental operations before actually using the BCI. To overcome this problem, it has been proposed to obtain BCI applications with very large training data bases, e.g. for multi-subject classification. Multi-subject classification has one more advantage --- it solves the problem of DNN long training times. Instead, a universal BCI system can be trained only once and then just applied to a new dataset from a new user without any additional training.~\cite{Lotte_2018} 

Guess the number (GTN) is a simple P300 event-related potential (ERP) BCI experiment. Its aim is to ask the measured participant to pick a number between 1 and 9. Then, he or she is exposed to corresponding visual stimuli. The P300 waveform is expected following the selected (target) number. During the measurement, experimenters try to guess the selected number based on manual evaluation of average ERPs associated with each number. Finally, both the numbers thought and guesses of the experimenters are recorded as metadata. 250 school-age children participated in the experiments that were carried out in elementary and secondary schools in the Czech Republic. Only three EEG channels (Fz, Cz, Pz) were recorded to decrease preparation time. Nevertheless, to the author's best knowledge, this is the largest P300 BCI dataset available so far.~\cite{pmid28350376} 

The main aim of this paper is to evaluate one of the deep learning models, convolutional neural networks (CNN) for classification of P300 BCI data. Unlike most related studies, multi-subject classification was performed with the future goal of developing a universal BCI. Two state-of-the art BCI classifiers were used as baseline to minimize the risk of biased comparison. To avoid overtraining, cross-validation and final testing using a previously unused part of the dataset were performed. Another aim of this manuscript is to evaluate some CNN parameters in this application.

\subsection{State-of-the-art}
Although various BCI algorithms have been evaluated and published in recent decades, there is still no feature extraction or machine learning algorithm clearly established as state-of-the-art. However, several studies have focused on reviews and comparisons with partly consistent results. In~\cite{Krusienski_2006}, a comparison of several classifiers (Pearson's correlation method, Fisher's linear discriminant analysis (LDA), stepwise linear discriminant analysis (SWLDA), linear support-vector machine (SVM), and Gaussian kernel support vector machine (nSVM)) was performed on 8 healthy subjects. It was shown that SWLDA and LDA achieved the best overall performance. As originally proposed by Blankertz~\cite{Blankertz2011} and also confirmed in a recent review~\cite{Lotte_2018},  shrinkage LDA is another useful tool for BCI, particularly with small training datasets.  In~\cite{Manyakov:2011:CCM:2043294.2064930}, the authors demonstrated that LDA and Bayesian linear discriminant analysis (BLDA) were able to beat other classification algorithms.

Efforts to develop a universal multi-subject P300 BCI machine learning have been relatively rare in the literature. In~\cite{notraining_sameperformance_p300}, the authors developed a generic shrinkage LDA classifier using the training data of 18 subjects. The performance was evaluated with the data of 7 subjects.  It was concluded that generic classifier achieved comparable results regarding the effectiveness and efficiency as personalized classifiers.  

\section{Methods}

\subsection{Data acquisition}
The data described in detail and accessible in~\cite{pmid28350376} were used in subsequent experiments. The measurements were taken between 8 am and 3 pm. Unfortunately, the environment was usually quite noisy since many children and also many electrical devices were present in the room at the same time. However, in any case there were no people standing or moving behind the monitor or in the close proximity of the measured participant. 

The participants were stimulated with numbers between 1 and 9 flashing on the monitor in random order. The numbers were white on the black background. The inter-stimulus interval was set to 1500 ms. The following hardware devices were used: the BrainVision standard V-Amp amplifier, standard small or medium 10/20 EEG cap, monitor for presenting the numbers, and two notebooks necessary to run stimulation and recording software applications. The reference electrode was placed at the root of the nose  and the ground electrode was placed on the ear. To speed up the guessing task, only three electrodes, Fz, Cz and Pz, were active. The stimulation protocol was developed and run using the Presentation software tool produced by Neurobehavioral Systems, Inc. The BrainVision Recorder was used for recording raw EEG data. 

The participants were school-age children and teenagers (aged between 7 and 17; average age 12.9), 138 males and 112 females. All participants and their parents were informed about the programme of the day and the experiments carried out. All participants took part in the experiment voluntarily. The gender, age, and laterality of the participants were collected. No personal or sensitive data were recorded. 

\subsection{Preprocessing and feature extraction}
\begin{figure}[h]
\includegraphics[scale=0.85]{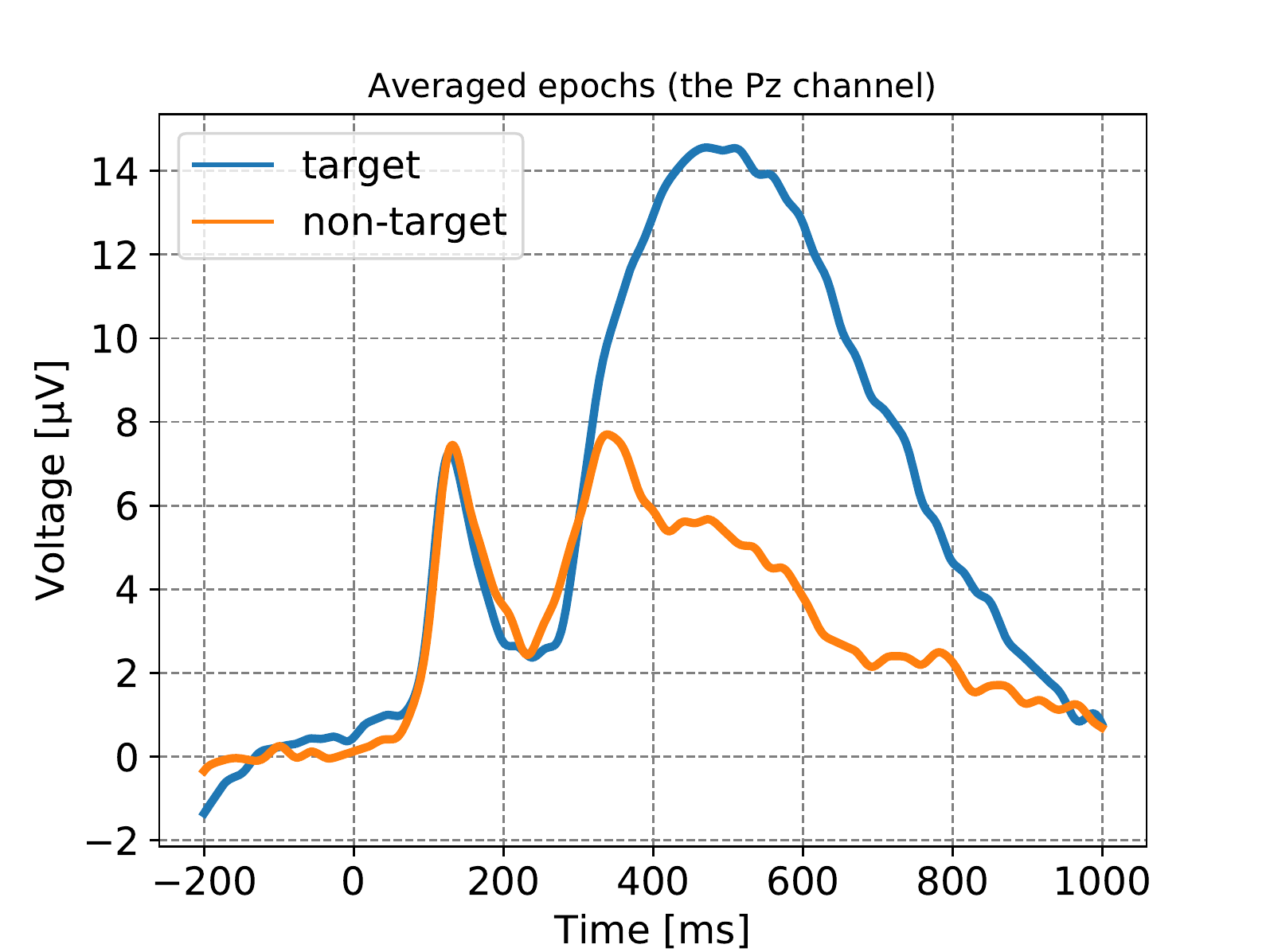}
\caption{Comparison of target and non-target epoch grand averages. As expected, there is a large P300 component following the target stimuli. Note that the P300 average latency is somewhat delayed compared to what is commonly reported in the literature~\cite{luck05introduction}.} 
\label{fig:epochavg}
\end{figure}

The data were preprocessed as follows:
\begin{enumerate}
    \item From each participant of the experiments, short parts of the signal (i.e. ERP trials, epochs) associated with two numbers displayed were extracted. One of them was the target (thought) number. Another one was randomly selected number out of the remaining stimuli between 1 and 9. Consequently, similar number of training examples for both classification classes (target, non-target) was extracted. The extracted epochs were stored into a file (available in~\cite{DVN/G9RRLN_2019}).
    \item For epoch extraction, intervals between 200 ms prestimulus and 1000 ms poststimulus were used. The prestimulus interval between -200 ms and 0 ms was used for baseline correction, i.e. computing average of this period and subtracting it from the data. Thus given the sampling frequency of 1 kHz, 11532 x 3 x 1200 (number of epochs x number of EEG channels x number of samples) data matrix was produced.
    \item To skip severely damaged epochs, especially caused by eye blinks or bad channels, amplitude threshold was set to 100 $ \mu V$ according to common guidelines (such as in~\cite{luck05introduction}). Any epoch $x[c, t]$ with $c$ being the channel index and $t$ time was rejected if:
    \begin{equation}
        \max_{c, t}|x[c, t]| > 100
    \end{equation}
    With this procedure, 30.3~\% of epochs were rejected. In Fig.~\ref{fig:epochavg}, grand averages of accepted epochs (across all participants) are depicted.
\end{enumerate}

\paragraph{Feature extraction}
Many deep learning methods such as CNN are designed to avoid significant feature engineering~(\cite{DBLP:journals/corr/abs-1206-5538} and \cite{journals/corr/ZeilerF13}). On the other hand, linear classifiers usually perform better when the dimensionality of the original data matrix is reduced, and only the most significant features are extracted~\cite{Blankertz2011}. In the parameter optimization phase, state-of-the-art classifiers were used either with original data dimension, or after feature selection proposed in~\cite{Blankertz2011} to compare the performance. The feature extraction method was based on averaging time intervals of interest and merging these averages across all relevant EEG channels to get reduced spatio-temporal feature vectors (Windowed means feature extraction, WM). In line with recommendations for P300 BCIs, a priori time window was initially set between 300 ms and 500 ms after stimuli~\cite{Tan:2010:BIA:1855009}. This time window was further divided into 20 equal-sized time intervals in which amplitude averages were computed. Therefore, with three EEG channels, the dimensionality of feature vectors was reduced to 60. Finally, these feature vectors were scaled to zero mean and unit variance.

\subsection{Classification}
\label{subsec:Classification}
\begin{figure}[h]
\includegraphics[scale=0.8]{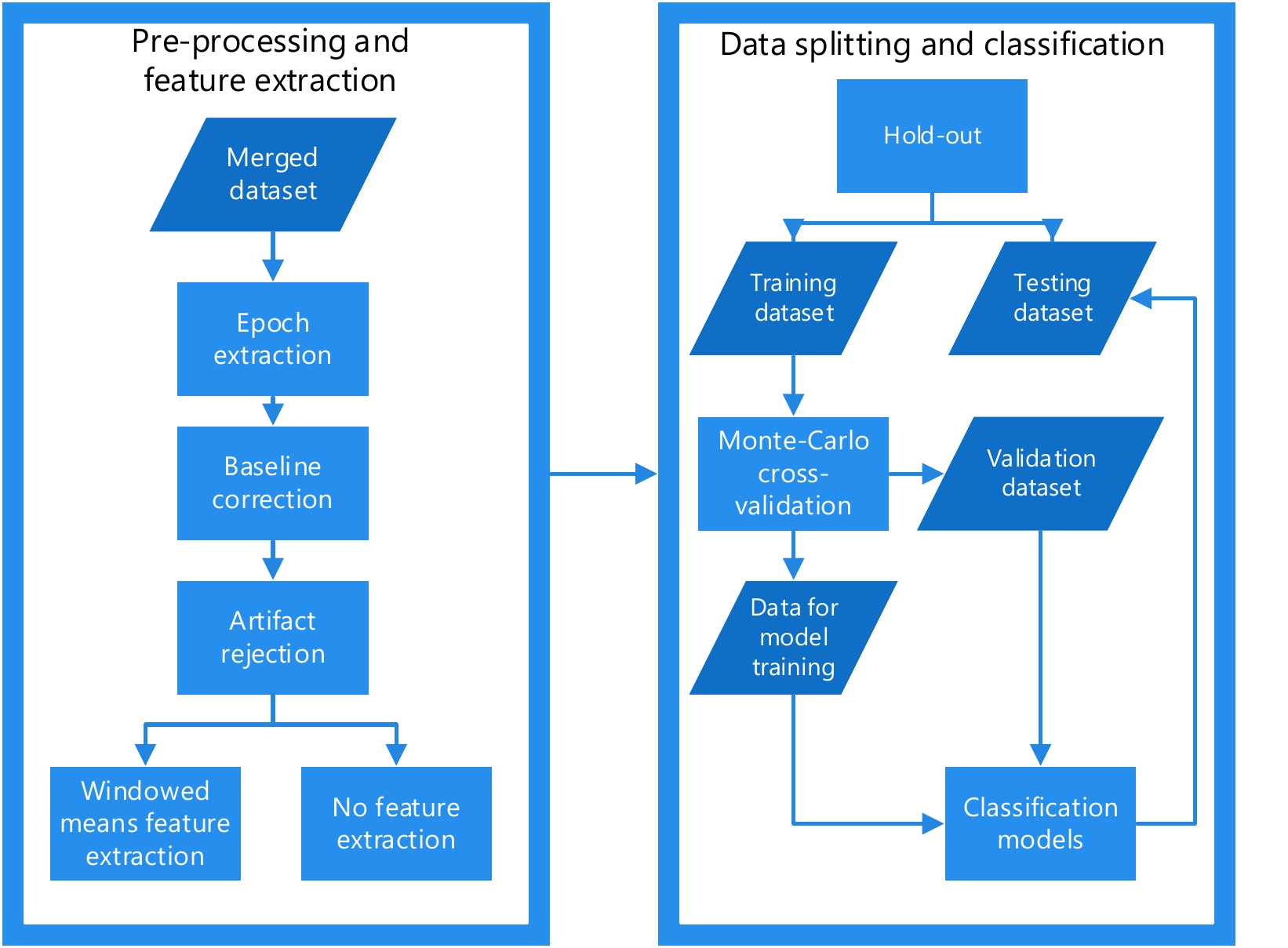}
\caption{Flowchart of preprocessing, feature extraction and data splitting applied.} 
\label{fig:flowchart}
\end{figure}

Fig. \ref{fig:flowchart} depicts procedures used to extract features and split the data for classification.

\paragraph{Data splitting}
Before classification, the data were randomly split into training (75~\%) and testing (25~\%) sets. Using the training set, 30 iterations of Monte-Carlo cross-validation (again 75:25 from the subset) were performed to optimize parameters. Results using the holdout testing set were computed in each cross-validation iteration and averaged at the end of the processing. No parameter decision was based on the holdout set. 

\paragraph{LDA}
State-of-the-art~\cite{Blankertz2011} LDA with eigenvalue decomposition used as the solver, and automatic shrinkage using the Ledoit-Wolf lemma~\cite{Ledoit110} was applied.

\paragraph{SVM}
The implementation was based on libsvm~\cite{CC01a}. Both recommendations in the literature~\cite{Fan:2008:LLL:1390681.1442794} and validation subsets were used to find the optimal parameters. Finally, penalty parameter $C$ was set to 1, the kernel cache was 500 MB, and degree of the polynomial kernel function was set to 3. One-vs-rest decision function of shape with the RBF kernel type and shrinking heuristics were used.

\begin{figure}[h]
\includegraphics[scale=0.45]{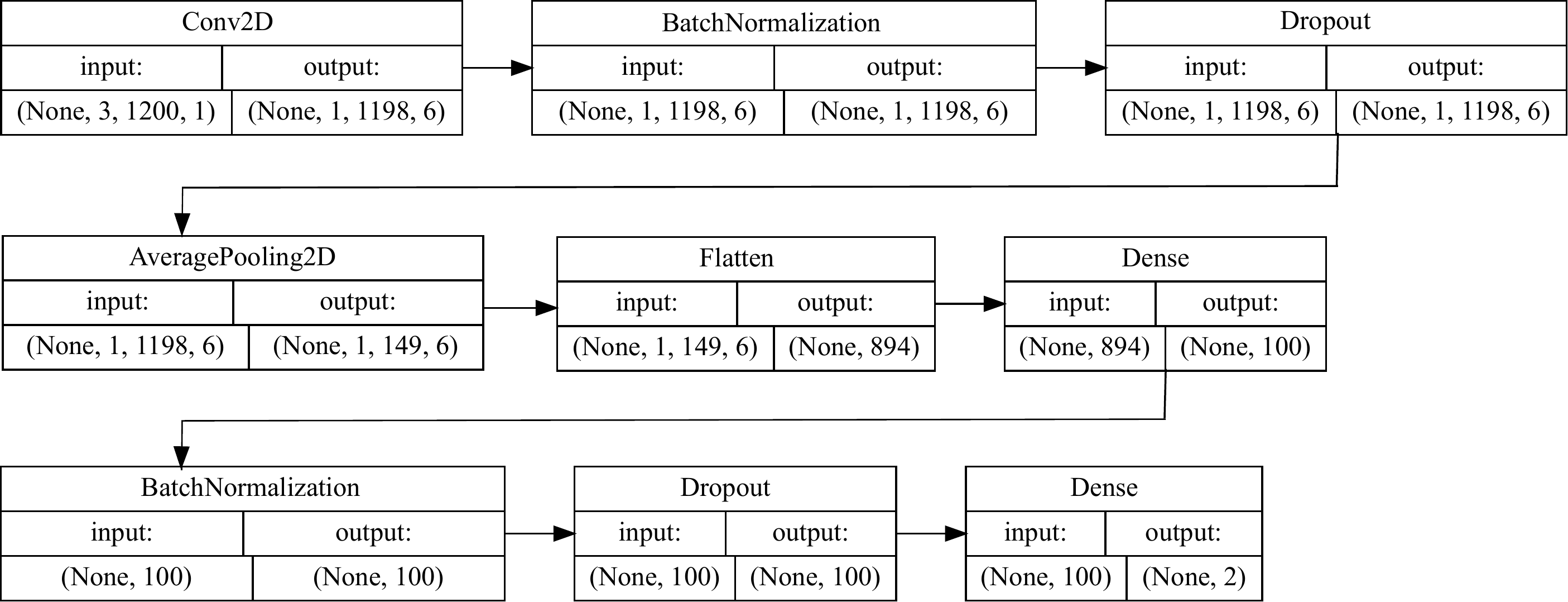}
\caption{Architecture of the convolutional neural network. There was one convolutional layer, one dense layer, and finally a softmax layer for binary classification (target / non-target). Batch normalization and dropout followed both the convolutional and dense layers. } 
\label{fig:model}
\end{figure}

\paragraph{CNN}
Convolutional neural networks were implemented in Keras \cite{chollet2015keras}. They were configured to maximize classification performance on the validation subsets. Its structure is depicted in Figure~\ref{fig:model}. Initially, after empirical parameter tuning based on cross-validation, the parameters were selected as follows:
\begin{itemize}
    \item The first convolutional layers had six 3~x~3 filters. The filter size was set to cover all three EEG channels. Both the second filter dimension and number of filters were tuned experimentally.
    \item In both cases, dropout was set to 0.5.
    \item The output of the convolutional layer was further downsampled by a factor of 8 using the average pooling layer.
    \item ELU activation function~\cite{Clevert2016FastAA} was used for both convolutional and dense layers as recommended in related literature~\cite{DBLP:journals/corr/SchirrmeisterSF17}. Compared to sigmoid function, ELU mitigates the vanishing gradient problem using the identity for positive values. Moreover, in contrast to rectified linear units (ReLU), ELUs have negative values which allow them to push mean unit activations closer to zero while ensuring a noise-robust deactivation state~\cite{Clevert2016FastAA}. The parameter $\alpha > 0$ was set to~1.
    \begin{equation*}
        f(x) = 
        \begin{cases}
            x & \text{if } x > 0  \\
            \alpha (e^{x} - 1) & \text{if } x \leq 0
        \end{cases}
    \end{equation*}
    
    \item Batch size was set to 16.
    \item Cross-entropy was used as the loss function.
    \item Adam~\cite{adam} optimizer was used for training because it is computationally efficient, has little memory requirements and is frequently used in the field~\cite{DBLP:journals/corr/abs-1901-05498}.
    \item The number of training epochs was set to 30.
    \item Early stopping with the patience parameter of 5 was used.
\end{itemize}

\section{Results}
\begin{figure}[h]
\includegraphics[scale=0.7]{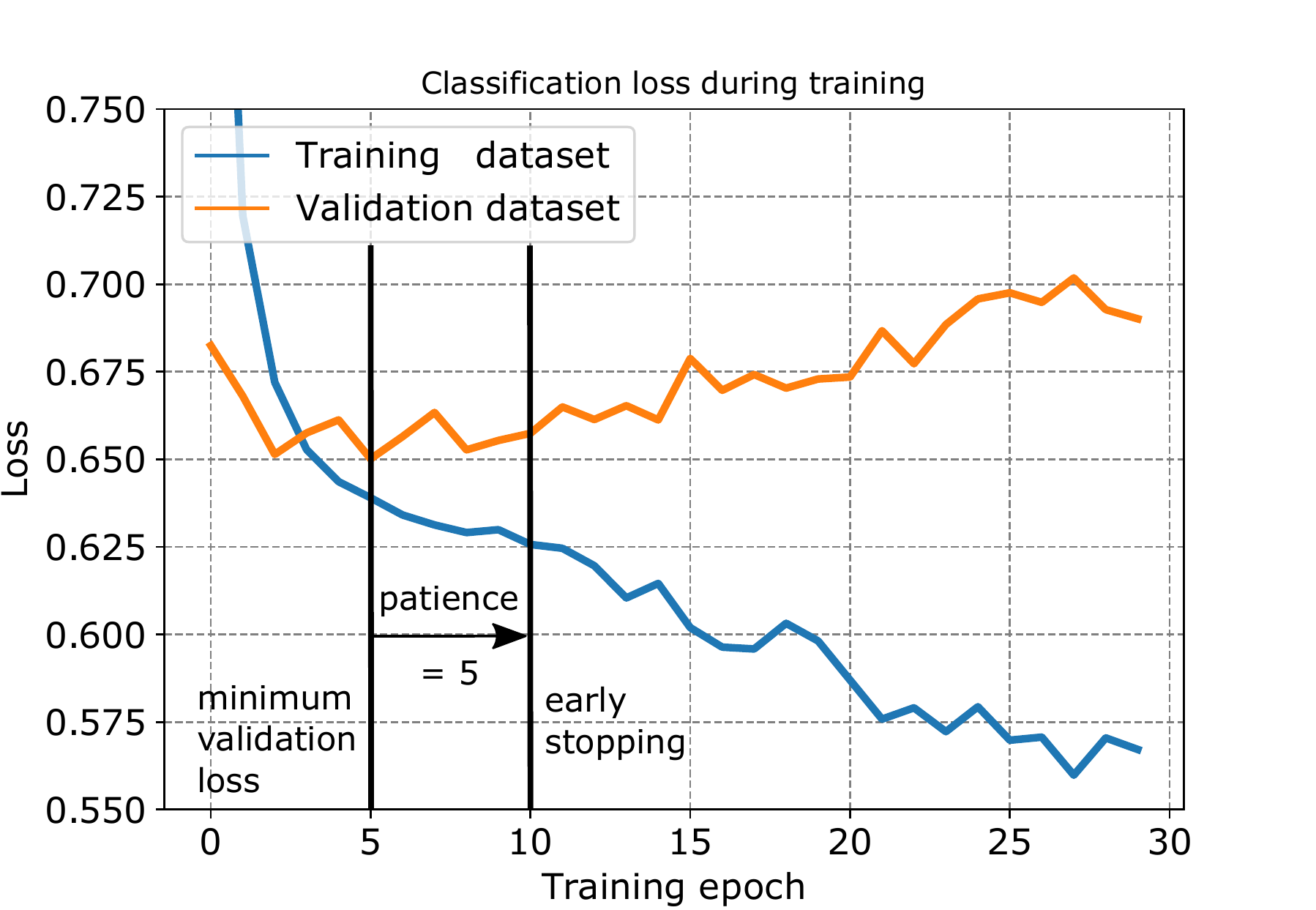}
\caption{Decrease of classification loss based on the baseline CNN architecture is shown. Although training loss kept declining throughout all 30 epochs, validation loss reached the minimum after only five epochs. Because the patience parameter was set to five, in this case, the training was stopped after 10 epochs. As seen from the growing difference between training and validation loss, further training would lead to substantial overtraining.} 
\label{fig:trainingloss}
\end{figure}

As mentioned above, cross-validation for hyperparameter estimation was followed by testing on a holdout set. Accuracy, precision, recall and AUC (Area under the ROC Curve) have been computed~\cite{hossin2015review}. In the validation phase, the aim was to reach the configuration yielding the highest accuracy while ensuring it is not at the expense of precision and recall. In Figure~\ref{fig:trainingloss}, an example of searching for an optimal configuration of CNN weights and biases based on the training and validation sets is shown. 

\subsection{Effect of parameter modifications on validation performance}
\label{sec:optimization}
\paragraph{Feature extraction for LDA and SVM}
Parameter optimization of the classifiers themselves has been discussed above. Additionally, different feature extraction settings were compared regarding the average classification results achieved during cross-validation. Results of the comparisons are depicted in Table~\ref{valfe}. Accuracy had an increasing trend when the time window got prolonged to 800 ms and 1000 ms. It can be speculated that the standard apriopri time window is not enough for capturing target to non-target differences when classifying children data that display a large variety in their P300 components. As expected, classification performance with WM features was slightly higher than for preprocessed epochs without feature extraction. Based on the results, both LDA and SVM configured as described above with the time window between 300 ms and 1000 ms were used in the testing phase.

\begin{table}
\caption{Average cross-validation classification results based on the feature extraction method with the LDA classifier configured as described in Subsection~\ref{subsec:Classification}. Averages from 30 repetitions and related sample standard deviations (in brackets) are reported. WM - Windowed means (time intervals relative to stimuli onsets in square brackets).}\label{valfe}
\begin{tabular}{|l|l|l|l|l|}
\hline
Feature extraction &  AUC & Accuracy & Precision & Recall\\
\hline
WM [300 - 500 ms]  & 59.56 \% (1.04)  & 59.54 \% (1.04)  & 59.48 \% (1.83) & 61.69 \% (2.08) \\
WM [300 - 800 ms]  & 60.94 \% (1.04)  & 60.93 \% (1.05)  & 60.75 \% (1.9)  & 63.38 \% (1.85) \\
\textbf{WM [300 - 1000 ms]} & 61.77 \% (0.9)   & \textbf{\underline{61.76}} \% (0.91)  & 61.45 \% (1.9) & 64.64 \% (1.48) \\
None               & 61.09 \% (1.13)  & 61.08 \% (1.13)  & 61.68 \% (1.67) & 59.90 \% (1.35) \\
 
\hline
\end{tabular}
\end{table}

\paragraph{CNN}
The neural network architecture described above was used as the starting point. However, some parameter modifications were explored regarding their effect on the validation classification results. The results are shown in Table~\ref{valresults}. Performance mostly displayed only small and insignificant changes with these parameter modifications. Consistently with~\cite{DBLP:journals/corr/SchirrmeisterSF17}, batch normalization led to slightly better accuracy. Moreover, the absence of batch normalization made the results less predictable and more fluctuating as can be seen in standard deviation of recall. Another clear decrease in performance was observed without dropout regularization. Finally, average pooling was better than max pooling for the validation data. Consequently, the initial configuration described in Subsection~\ref{subsec:Classification} was used for testing.

\begin{table}
\caption{Average cross-validation classification results based on the CNN parameter settings. Averages from 30 repetitions and related sample standard deviations (in brackets) are reported. CNN configuration described in Subsection~\ref{subsec:Classification} was used as the baseline model.}\label{valresults}
\begin{tabular}{|l|l|l|l|l|}
\hline
Changed parameter &  AUC & Accuracy & Precision & Recall\\
\hline
\textbf{None}              & 66.12  \% (0.68) &  \textbf{\underline{62.18}}  \% (0.94) &  62.76  \% (1.95) &  61.34  \% (2.63) \\
RELUs instead of ELUs      & 66.36  \% (0.62) &  61.85  \% (1.15) &  62.7  \% (2.19) &  60.1  \% (3.04) \\
Filter size (3, 30)        & 65.84  \% (0.49) &  61.95  \% (1.18) &  62.7  \% (2.1) &  60.5  \% (3.91) \\
12 conv. filters           & 66.31  \% (0.51) &  61.83  \% (1.1) &  62.3  \% (2.21) &  61.6  \% (3.08) \\
No batch normalization     & 65.99  \% (0.77) &  60.55  \% (1.52) &  61.02  \% (3.16) &  61.5  \% (7.21) \\
Dropout 0.2                & 67.67  \% (0.65) &  60.8  \% (1.49) &  61.33  \% (2.31) &  60.33  \% (4.0) \\
No dropout                 & 68.63  \% (1.11) &  59.49  \% (1.2) &  59.61  \% (1.93) &  60.7  \% (4.44) \\
Dense (150)                & 66.07  \% (0.8)  &  61.81  \% (0.95) &  62.33  \% (1.83) &  61.18  \% (2.49) \\    
Two dense l. (120-60)  & 65.72  \% (0.77) &  62.11  \% (0.9) &  63.14  \% (2.03) &  59.5  \% (2.55) \\
Max- instead of AvgPool & 64.23  \% (1.15) &  58.94  \% (1.94) &  60.22  \% (4.18) &  59.24  \% (13.76) \\
\hline
\end{tabular}
\end{table}

\begin{figure}[h]
\includegraphics[scale=0.7]{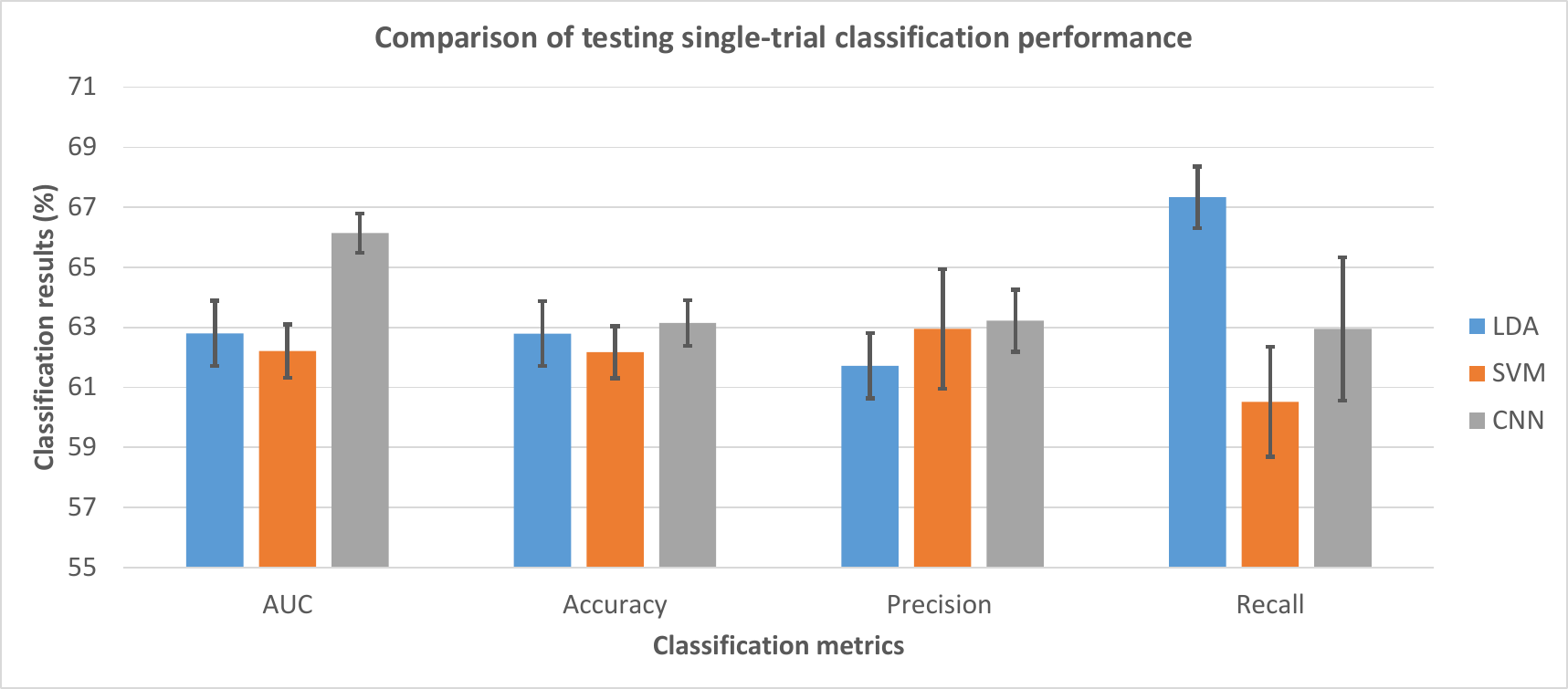}
\caption{Testing results for single trial classification (error bars show standard deviations).} 
\label{fig:results}
\end{figure}

\subsection{Testing results}
Based on the results in Subsection~\ref{sec:optimization}, both feature extraction method for LDA and SVM, and CNN configuration achieving the best average accuracy during cross-validation were selected for the testing phase. Figure~\ref{fig:results} shows the achieved results. All tested models achieved comparable classification results. LDA had the highest classification recall (around 67 \%). Single trial classification accuracy stayed within the range between 62~\% and 64~\%. 

\begin{figure}[h]
\includegraphics[scale=0.75]{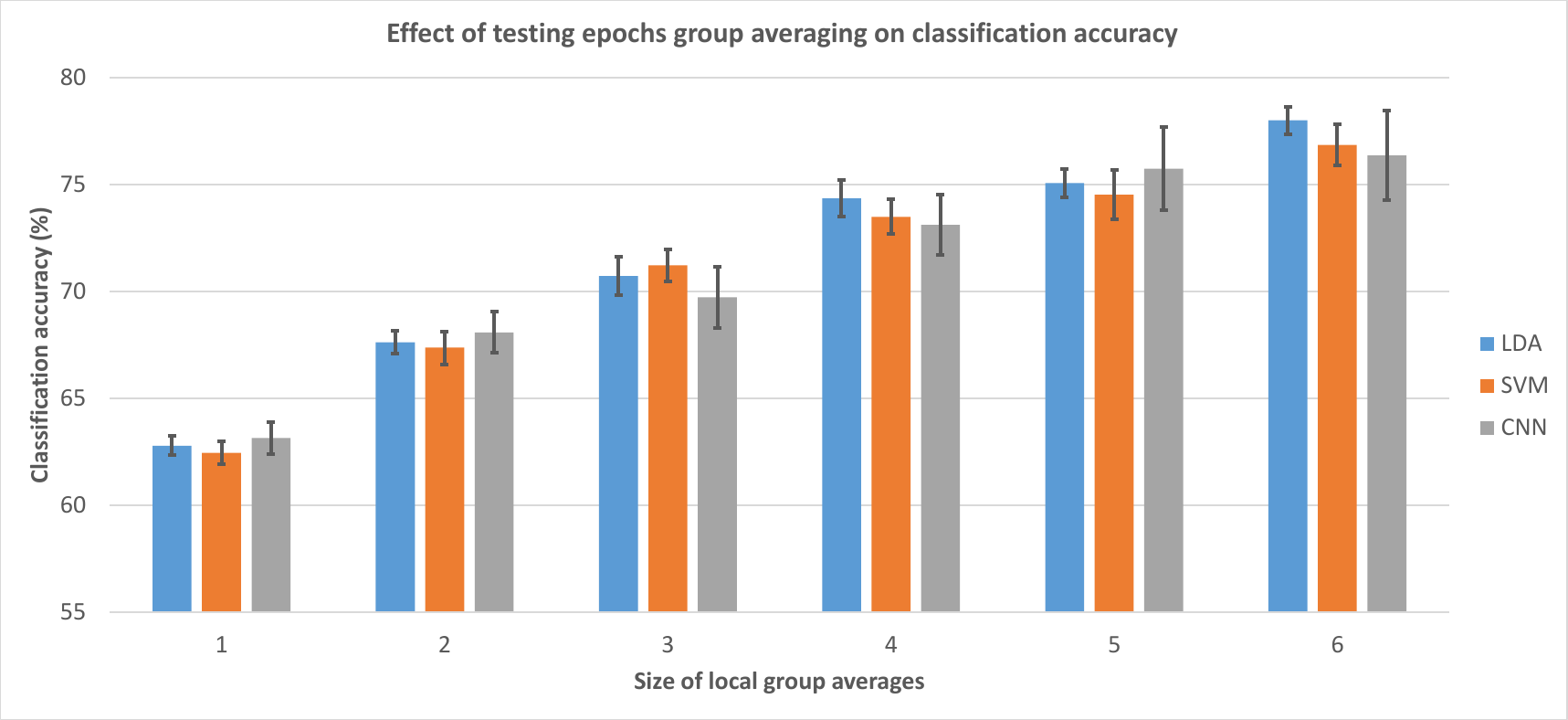}
\caption{Testing results when averaging neighboring epochs (error bars show standard deviations).} 
\label{fig:resultsavg}
\end{figure}

Averaging of epochs associated with the same markers is a standard ERP technique for increasing signal-to-noise ratio~\cite{luck05introduction}. When averaging, repeated ERPs including the P300 are amplified while continuous random EEG noise is suppressed. Because even in P300 BCIs, repeated stimulation is usually used to achieve good performance~\cite{pmid21067970}, it is worth exploring how once trained classifiers can generalize to averaged epochs. Therefore, consecutive groups of one to six neighboring epochs from the testing set were used instead of single trials. Fig.~\ref{fig:resultsavg} depicts the results achieved. With averaging, classification accuracy increased from original 61 - 64~\% up to 76 - 79~\%. There were no significant differences among classifiers, although CNN displayed slightly higher standard deviations.

\section{Discussion}
Single trial classification accuracy was between 62~\% and 64~\% for all tested classification models without significant differences. Similar results have been commonly reported in the literature. For example, in~\cite{6599576}, 65~\% single trial accuracy was achieved (using one to three EEG channels and personalized training data). In~\cite{DBLP:journals/corr/abs-1712-01977}, 40~\% to 66~\% classification accuracy was reported, highly dependent on the tested subject. Comparably, this manuscript achieved similar performance for a large multi-subject dataset of school-age children. 

On single trial level, CNN achieved comparable performance to both LDA and SVM. Similar performance was also achieved when applying averaged testing epochs. However, CNN seemed slightly less stable and more dependent on training/validation split as can be seen in standard deviations.

\begin{figure}[!h]
\centering
\includegraphics[scale=0.4]{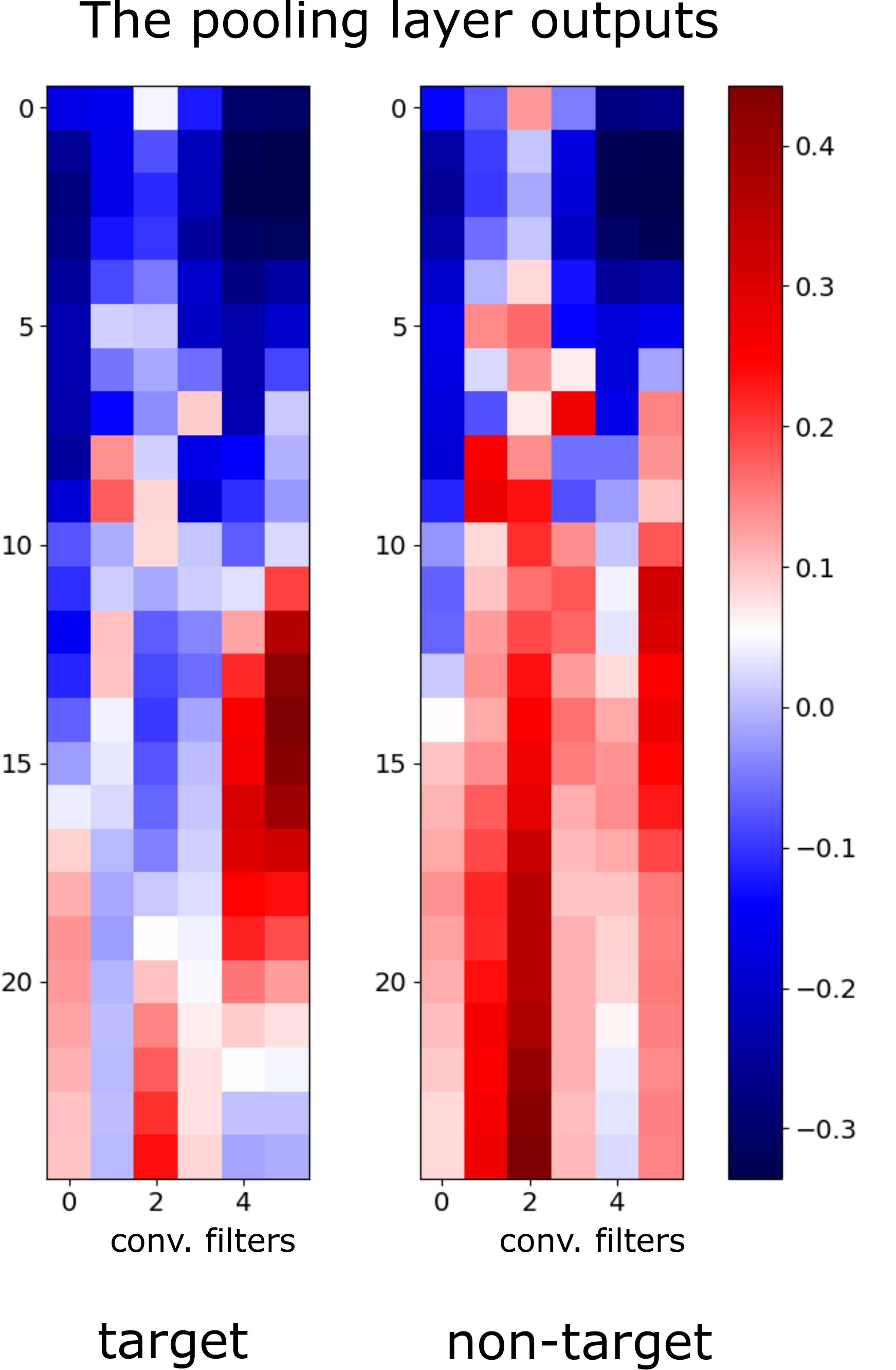}
\caption{Average outputs of the 4th (pooling) layer are depicted after the CNN was exposed to all target / non-target patterns. X-axis corresponds to indices of convolutional filters (six in total). Y-axis is the output of convolution originally corresponding to time information, after average pooling further downsampled by a factor of 6. There is a clear difference in outputs, mainly in the bottom part of the maps. However, many outputs seem independent of classification labels, poorly contributing to CNN discrimination abilities.} 
\label{fig:hiddenoutputs}
\end{figure}

Consistently with related deep learning literature~\cite{DBLP:journals/corr/SchirrmeisterSF17}, a combination of ELUs, dropout and batched normalization were beneficial for classification performance. Unlike many image classification applications, average pooling was better than max pooling, perhaps because it is not associated with data loss. Even less prominent features may contribute to classifier discriminative abilities. To further verify how the CNN was able to classify between targets and non-targets, the network was exposed to all target, or all non-target patterns. Average hidden layer outputs (the 4th average pooling layer used as an example) across these conditions were calculated and shown in Fig.~\ref{fig:hiddenoutputs}. There is a clear difference between some CNN outputs although the most remain stable across both conditions.

In our previous work~\cite{icaisc2017}, we applied stacked autoencoders (SAE) to the same GTN dataset. In contrast with the current work, manual feature extraction using discrete wavelet transform was performed. Instead of single trial classification, success rate of detecting the number thought based on multiple single trial classification results was computed. Maximum success rate on the testing dataset was 79.4~\% for SAE, 75.6~\% for LDA and 73.7~\% for SVM. It seems that while SAE combined with traditional feature engineering and involving multiple trials per marker can outperform linear classifiers, the same benefits cannot be repeated when applying CNN to single trial classification of raw EEG data.

Computational efficiency is another important factor to consider when applying the methods in online BCI systems. Experimental comparison was performed with Intel Core i7-7700K, four cores, 4.2 GHz, 64 GB RAM and NVIDIA GeForce GTX 1050 Ti GPU. CNN took 46 seconds to train on CPU and 26 s to train on GPU. Both LDA and SVM were much faster to train, with 300 ms and 1600 ms, respectively. However, training times were not critical in the presented experiment since any universal classifier needs to be trained just once and not with every new BCI user. Testing times were calculated relative to one processed feature vector and were low enough for all classifiers (CNN took 0.3 ms to classify one pattern on CPU and 0.1 ms on GPU, LDA took 0.1 ms and SVM 0.2 ms). It can be concluded that all tested algorithms can be used in online BCIs. Neural networks are slower to train and this could be a problem for personalised BCIs, retrained with each new user.

There are several limitations of the reported experiments. As a noise suppression procedure, severely damaged epochs (with amplitude exceeding $\pm 100 \mu V$ when compared to baseline) were rejected before further processing. While epoch rejection is beneficial for classification accuracy, on the other hand, it would also lead to lower bit-rates when used in on-line P300 BCI systems because of data loss. Artifact correction methods based on Independent Component Analysis were not feasible because of the low number of EEG channels (three). Moreover, the low number of EEG channels could have a detrimental effect on classification performance because of limited spatial information provided on the input.  Another possible limitation was that there might be an architecture of CNN that would lead to better classification performance and had not been discovered by the author. However, several manipulations of CNN parameters were tested using cross-validation, including adding a new dense layer, with only very modest changes in validation classification accuracy.

Recent review of EEG and DNNs~\cite{DBLP:journals/corr/abs-1901-05498} studies reported the median gain in accuracy of DNNs over traditional baselines to be 5.4~\%. It also revealed significant challenges in the field. Low number of training examples is a common complaint especially for event-related data that contain the relevant information in time domain. In this case, only a small fraction of continuous EEG measurement near the onset of trials can be used and strategies such as overlapping time windows to obtain more examples in frequency domain are not feasible. In the current study, 11,532 epochs were used which is below mean number of examples (251,532) and medium number of examples (14,000) in the reviewed papers~\cite{DBLP:journals/corr/abs-1901-05498}. Strategies such as data augmentation can be considered to increase the number of training examples to be sufficient for DNNs. Moreover, half of the studies~\cite{DBLP:journals/corr/abs-1901-05498} used between 8 and 62 EEG channels. Adding more channels to Fz, Cz and Pz could increase spatial resolution and accuracy but would also increase preparation time and the participant's discomfort. In future work, more on the effect of number of EEG channels on the P300 classification accuracy can be investigated. Furthermore, soft or hard thresholding based on discrete wavelet transform can be considered for noise cancellation~\cite{pmid23142653}. Another line of research would be to propose different deep learning models for the same classification task, with extensive parameter grid search, or genetic algorithms. Based on the recent review of the field~\cite{DBLP:journals/corr/abs-1905-04149}, frequently cited and promising networks include Recurrent Neural Networks, especially Long short-term memory (LSTM). Moreover, a CNN layer to capture spatial patterns can be followed by a LSTM layer for temporal feature extraction~\cite{DBLP:journals/corr/abs-1905-04149}.

\section{Conclusion}
The aims of the presented experiments were to compare CNN with baseline classifiers (LDA, SVM) using a large multi-subject P300 dataset. CNN was applied to raw ERP epochs (with the dimensionality of 3~x~1200). Baseline classifiers were applied to windowed means features (with the dimensionality of 60). Empirical parameter optimization was performed using cross-validation and classifiers were tested on a holdout set. Various CNN parameters are discussed. Single trial classification accuracy was between 62~\% and 64~\% for all tested models with CNN able to match but not outperform its competitors. When the trained models were applied to averaged trials in the testing phase, accuracy increased up to 76 - 79~\%. Achieved accuracy is comparable with state-of-the-art despite using a multi-subject dataset from 250 children. Potential explanation of the results are discussed. Based on the results, LDA and SVM with state-of-the-art feature extraction still seem to be a good choice for P300 classification, especially with relatively small training datasets. CNN might need more spatial information in the data (by means of more channels) to better understand the patterns. Alternatively, the dataset was not large enough for CNN to prove its benefits and e.g. data augmentation techniques could help to overcome this obstacle. Both the preprocessed data~\cite{DVN/G9RRLN_2019} and Python codes~\cite{VarekaBitbucket} are available to ensure reproducibility of the experiments. 

\section*{Acknowledgement}
This publication was supported by the project LO1506 of the Czech Ministry of Education, Youth and Sports under the program NPU I.

\clearpage 

%
%
%
\bibliographystyle{splncs04}
\bibliography{bibliography}
\end{document}